\documentstyle[12pt,aaspp4,amstex]{article} 
\def\hh{\tiny {\mbox{HI}}}
\def\hhh{\tiny {\mbox{HII}}}
\def\he{\tiny {\mbox{HeI}}}
\def\heh{\tiny {\mbox{HeII}}}
\def\hehh{\tiny {\mbox{HeIII}}}	

\def\ltsima{$\; \buildrel < \over \sim \;$}
\def\simlt{\lower.5ex\hbox{\ltsima}}
\def\gtsima{$\; \buildrel > \over \sim \;$}
\def\simgt{\lower.5ex\hbox{\gtsima}}

\begin{document}

\title{The Nature of the Ionising Background at ${\bf z\approx 2.5 - 5}$}

\author{Aaron Sokasian\footnote{asokasia@@cfa.harvard.edu}}
\author{Tom Abel\footnote{hi@@tomabel.com} and Lars
Hernquist\footnote{lars@@cfa.harvard.edu}} \affil{Department of
Astronomy, Harvard University, Cambridge, MA 02138}
\authoremail{asokasia@cfa.harvard.edu}

\begin{abstract}

Using radiative transfer calculations and cosmological simulations of
structure formation, we study constraints that can be placed on the
nature of the cosmic ultraviolet (UV) background in the redshift
interval $2.5\simlt z \simlt 5$.  Our approach makes use of observational
estimates of the opacities of hydrogen and singly ionised helium in
the intergalactic medium during this epoch.  In particular,
we model the reionisation of He{\small II} by sources of hard
ultraviolet radiation, i.e. quasars, and infer values for our
parameterisation of this population from observational estimates of
the opacity of the He{\small II} Lyman-alpha forest.  Next, we
estimate the photoionisation rate of H{\small I} from these sources
and find that their contribution to the ionising background is
insufficient to account for the measured opacity of the H{\small I}
Lyman-alpha forest at a redshift $z\sim 3$.  This motivates us to
include a soft, stellar component to the ionising background to boost
the hydrogen photoionisation rate, but which has a negligible impact
on the He{\small II} opacity.

In order to simultaneously match observational estimates of the
H{\small I} and He{\small II} opacities, we find that galaxies and
quasars must contribute about equally to the ionising background in
H{\small I} at $z\simeq 3$.  Moreover, our analysis requires the
stellar component to rise for $z > 3$ to compensate for the
declining contribution from bright quasars at higher redshift.  This
inference is consistent with some observational and theoretical
estimates of the evolution of the cosmic star formation rate.  The
increasing dominance of the stellar component towards high redshift
leads to a progressive softening of the UV background, as suggested by
observations of metal line absorption.  In the absence of additional
sources of ionising radiation, such as mini-quasars or weak active
galactic nuclei, our results, extrapolated to $z > 5$, suggest that
hydrogen reionisation at $z\sim 6$ mostly likely occurred through the
action of stellar radiation.

\end{abstract}

\keywords{radiative transfer -- diffuse radiation -- intergalactic
medium -- galaxies: quasars}

\section{INTRODUCTION}

Determining the relative contributions of different sources to the
cosmic ultraviolet background is essential for understanding the
evolution of the intergalactic medium (IGM).  In particular, this
metagalactic radiation field is believed to have reionised hydrogen at
$z\sim 6$ (e.g. Becker et al. 2001) and helium slightly later,
although the epoch of helium reionisation has yet to be determined
observationally (for a discussion, see e.g. Sokasian et al. 2002).
Evidence from measured temperature changes, optical depth variations,
and evolution in the relative abundances of metal line absorbers
strongly suggests that most intergalactic helium became fully ionised
at redshifts close to $\sim 3.2$ (e.g.  Davidsen et al. 1996, Jakobsen
et al. 1994, Kriss et al. 2001, Reimers et al. 1997, Songaila 1998,
Ricotti, Gnedin \& Shull 1999, Theuns et
al. 2002a, Theuns et al. 2002b, Bernardi et al. 2002 and references
therein).

An important probe of the physical state of the intergalactic medium
is provided by bright objects at great distances, such as quasars.
For example, it is now believed that absorption by diffuse,
cosmologically distributed gas is responsible for the hydrogen Lyman
alpha forest (e.g. Cen et al. 1994; Zhang et al. 1995; Hernquist et
al.  1996).  Similarly, Ly$\alpha$ absorption by He {\small II} along
a line of sight to a distant quasar probes gas in the intervening IGM
at even lower overdensities (Croft et al. 1997), characteristic of
much of the baryonic matter in the Universe (e.g. Dav\'e et al. 2001;
Croft et al. 2001).  At a given redshift, the number and strengths of
these spectral features is sensitive to the local density of absorbing
atoms, which in turn depends on the gas density, cosmological
parameters, and the intensity of the ionising background. In fact,
much of the interpretation of spectroscopic observations of high
redshift quasars relies strongly on this simple picture of the Lyman
alpha forest.

Specifically, given a model for the formation of large-scale
structure, the number of lines detected in the Lyman-$\alpha$ forest
as a function of redshift directly constrains the evolution and
spectral properties of the radiation field.  In a recent study, Kim,
Cristiani, \& D'Odorico (2001) showed that the number of lines per
unit redshift, $dN/dz$, with column densities in the interval
$N_{\text{H {\tiny I}}}=10^{13.64 - 16}$ decreases continuously from
$z\sim 4$ to $z\sim 1.5$ according to $dN/dz\propto
(1+z)^{2.19\pm0.27}$.  Combined with the results of Weymann et
al. (1998), who find a much flatter distribution for $dN/dz\propto
(1+z)^{0.16\pm0.16}$ at $z<1$ , it appears that the line number
density of the Ly$\alpha$ forest is well described by a double
power-law with a break at $z\sim 1$.  These results suggest that the
evolution of the forest above $z>1.5$ is governed mainly by Hubble
expansion and that there is little change in the ionising background
until the break occurring at $z\sim 1$.

The location of the observed break, however, is inconsistent with
theoretical predictions derived from numerical simulations.  In
particular, studies of the Lyman-$\alpha$ forest carried out by Dav\'e
et al. (1999) and Machacek et al. (2000) predict a break in the double
power-law occurring near $z\sim 1.8$.  While these simulations have
provided a successful general description of the evolution of the
Lyman-$\alpha$ forest, their apparent inability to match the location
of the break indicates that the underlying assumptions regarding the
form of the UV background may be incorrect. More specifically, these
simulations assume a quasar (QSO) type source population mainly
responsible for producing the radiation field. Since the emissivity of
quasars is known to fall off steeply below $z \sim 2$, so
would their contribution to the UV background, thereby producing a
break in $dN/dz$ around this redshift.  One way to reconcile the
inconsistency between the simulations and observations is to appeal to
other types of sources to maintain the intensity of the UV background
at a relatively high level until $z\sim 1$.

Recently, Bianchi et al. (2001) have explored the possibility that
galaxies might provide this additional contribution to the radiation
field.  In particular, they derived the H {\small I} ionising
background resulting from the integrated contribution of quasars and
galaxies, taking into account the opacity of the intervening IGM.  The
quasar emissivity was derived from fits to an empirical luminosity
function, while a stellar population synthesis model and a cosmic
star-formation history from UV observations were used to estimate the
galaxy emissivity.  They found that the break at $z\sim 1$ implied by
the Kim et al. (2001) analysis can be understood if the contribution
from galaxies is comparable to or larger than that of quasars.  This
is consistent with other determinations of the galactic component of
the background (Giallongo, Fontana, \& Madau 1997; Devriendt et
al. 1998; Shull et al. 1999; Steidel, Pettini, \& Adelberger 2001).  A
significant contribution to the radiation field from galaxies would
imply a considerable softening of its spectrum compared to previous
models which included only quasars as the dominant source of the
ionising metagalactic flux.

In this paper, we shift the focus to higher redshifts to see whether
including an additional component from galaxies together with a
realistic quasar model is capable of producing the required ionising
intensity to match observations of H {\small I} photoionisation rates
in the redshift range $2.5<z<5$.  Our method involves combining
numerical and empirical results on the reionisation of singly ionised
helium (He {\small II}) by quasars as a tool for estimating the
additional contribution from galaxies to the background which would be
necessary to simultaneously match results derived from the H {\small
I} Ly$\alpha$ forest.  In particular, we select a quasar model based
on the study conducted in our earlier paper (Sokasian, Abel \&
Hernquist 2002; hereafter SAH), where we applied a numerical method
to study the 3D reionisation of He {\small II} by quasars.
The adopted model is then used to calculate the contribution to the H
{\small I} background from quasars only.  We can then estimate the
required contribution from galaxies to match the H {\small I}
photoionisation rates measured by Rauch et al. (1997) in the redshift
range $2.5<z<5$.  Consequently, this approach allows us to make
estimates of the amplitude and evolution of this component.

This paper is organised as follows.  In Sections 2, 3, and 4 we
describe our approach for calculating the emissivities of quasars,
galaxies, and recombination radiation from the IGM, respectively.  The
procedure for determining the ionising background for hydrogen given
these emissivities is presented in Section 5.  Results of our analysis
are discussed in Section 6 and conclusions are given in Section 7.

\section{METHOD}

In this section we describe our motivation for adopting a specific QSO
model for the purpose of deriving the associated contribution to the H
{\small I} ionising background.  This component may be estimated from
a QSO luminosity function (LF).  However, this requires making a
number of assumptions concerning the emission from the sources and
how easily this radiation can escape into the IGM.  Instead, we will
constrain the quasar component of the ionising background by appealing
to numerical modeling of He {\small II} reionisation along with
observational estimates of the evolution of the He {\small II} opacity.

To this end, we compile a list of quasar type sources which were
selectively extracted from a cosmological simulation according to a
QSO LF and a set of characteristic source parameters.  This approach
provides us with a source list which is directly applicable as input
for a cosmological radiative transfer simulation designed to examine
the He {\small II} reionisation process (as in SAH).  The advantage of
such an approach is that it provides us with a way of choosing the
most successful model for our analysis based on a comparative study
between the numerical and empirical results for the He {\small II}
opacities measured in quasar spectra.

The numerical scheme used to calculate the 3D reionisation of He
{\small II} by quasars is described in Sokasian, Abel, \& Hernquist
(2001).  In SAH, we used this approach to explore the parameter space
associated with the characteristics of the sources and studied how
they influenced global properties of the reionisation process.
Comparisons with observational results were made possible by
extracting synthetic spectra from the simulations.  There our aim was
twofold: to develop an understanding of the sensitivity of the
reionisation process to source properties and to examine the
predictions of the different models in light of recent observational
results.

The cosmological simulation we used in SAH was based on a smoothed
particle hydrodynamics (SPH) treatment, computed using the parallel
TreeSPH code 
GADGET developed by Springel, Yoshida \& White (2001).  The
particular cosmology we examine is a $\Lambda$CDM model with
$\Omega_{\rm b}=0.04$, $\Omega_{\rm DM}=0.26$, $\Omega_{\Lambda}=0.70$, and
$h=0.67$ (see, e.g., Springel, White \& Hernquist 2001).  The
simulation uses $224^{3}$ SPH particles and $224^{3}$ dark matter
particles in a $67.0 \ \text{Mpc}/h$ comoving periodic
box, resulting in mass
resolutions of $2.970\times 10^8 \ M_{\odot}/h$ and $1.970\times 10^9
\ M_{\odot}/h$ in the gas and dark matter components, respectively.
The gas can cool radiatively to high overdensity (e.g.  Katz, Weinberg
\& Hernquist 1996) and is photoionised by a diffuse radiation field
which is assumed to be of the form advocated by Haardt \& Madau (1996;
see also, Dav\'e et al. 1999).  When sources are included in our
treatment of helium reionisation, the ionisation state of the helium
is recalculated, ignoring the diffuse background that was included in
the hydrodynamical simulation (see Sokasian et al. 2001 for details).

\subsection{The QSO model}

In SAH, QSO models were differentiated from one another based on their
respective values for the free parameters associated with the source
selection algorithm.  The full details of our scheme are described in
\S 2 and \S 3 of SAH. The basic procedure involves identifying dense
clumps of gas in the cosmological simulations which represent
plausible quasar sites, and adopting a prescription for selecting a
subset of these objects as actual sources according to an empirical
quasar luminosity function.  In our analysis, we choose the double
power-law form of the quasar luminosity function presented by Boyle
et al. (1988) using the open-universe fitting formulae from Pei (1995)
for the B-band (4400 \AA \ rest-wavelength) LF of observed quasars,
with a rescaling of luminosities and volume elements for our
$\Lambda$CDM cosmology.  Our selection algorithm requires us to adopt
an evolving mass-to-light ratio, $\xi(z)$, which scales with $z$ as the
break luminosity $L_z$ inherent in the LF.

Once a source with mass, $M$, has been selected, it is assigned a
B-band luminosity, $L_{B}=M/\xi(z)$ (in ergs $\mbox{s}^{-1}$).
Along with an assumed spectral form, this luminosity is then used to
compute the amount of ionising flux that will be generated while the
source is active.  For all sources, we assume a piece-wise power-law form
for the spectral energy distribution (SED),
\begin{equation}
L(\nu)\propto
\begin{cases}
\nu^{-0.3} & (2500\ \mbox{\AA}<\lambda<4400\ \mbox{\AA});\\ \nu^{-0.8}
& (1050\ \mbox{\AA}<\lambda<2500\ \mbox{\AA});\\ \nu^{-\alpha_{\rm QSO}} &
(\lambda<1050\ \mbox{\AA}),
\end{cases}
\end{equation}
where a choice of $\alpha_{\rm QSO}=1.8$ corresponds to the SED proposed 
by Madau, Haardt, \& Rees (1999) based on the rest-frame optical and 
UV spectra of Francis et al. (1991) and Sargent, Steidel, \& Boksenberg 
(1989), and the EUV spectra of radio-quiet quasars (Zheng et al. 1998).

The entire selection process, including the assigning of intensities,
introduces five free parameters associated with source
characteristics.  They are: (1) a universal source lifetime,
$T_{\rm life}$, (2) a minimum mass, $M_{\rm min}$, (3) a minimum luminosity
at $z=0$, $L_{\rm min,0}$, (4) an angle specifying the beaming of the
bi-polar radiation, $\beta$, and (5) a tail-end spectral index,
$\alpha_{\rm QSO}$, in the regime $\lambda< 1050$ \AA. In SAH, we computed
and analysed six models with different sets of values for the free
parameters.  Below, we discuss the parameter choices for Models 1 and
5 which represent our fiducial and {\it best fit} models,
respectively.  Table 1 lists the corresponding parameter choices.

In Model 1, our fiducial case, we adopted a widely quoted value for
$L_{\rm min,0}$ based on the results of Cheng et al. (1985) who show that
the LF of Seyfert galaxies (which is well correlated with that of
optically selected quasars at $M_B=-23$) exhibits some evidence of
leveling off by $M_B\simeq -18.5$ or $L_{\rm min}\simeq 3.91\times 10^{9}
\ L_{B,\odot}$ at $z=0$. The tail-end spectral index parameter for
this model was chosen to be $\alpha=1.8$, making the SED in this model
identical to the one advocated by Madau, Haardt, \& Rees (1999). In
Model 5, we examined the effect of reducing the ionising emissivity
from the sources. This was accomplished by increasing $L_{\rm min,0}$ by a
factor of six and by steepening the tail-end spectral index to
$\alpha=1.9$. In both models, the sources are assumed to radiate their
flux isotropically ($\beta=\pi$) and to have a minimum mass cut off of
$M_{\rm min}=1.80 \times 10^{10}\ M_{\odot}$, which produces good agreement
with the B-band emissivity predicted by the LF and which provides a
realistic mass function given the level of resolution of the
simulation at this mass limit (see SAH for further details).  The value
of the mass cut off is also consistent with the assumption that the
sources are galaxy type objects acting as quasar hosts.

Figure 1 shows the redshift evolution of the effective mean optical
depth for He {\small II} absorption derived from both models.  The
mean optical depth is defined as $\bar{\tau}_{\mbox{\tiny{He
II}}}\equiv-\log_e \langle T \rangle$, where $T$ is the transmittance
extracted from the synthetic spectra from each model. The average is
performed over 500 lines of sight within 35 wavelength bins of width
$\Delta \lambda=6.57$ \AA. Hatched regions represent the optical depth
derived from the simulations at the 90\% confidence level with the
dashed lines indicating mean values.  For comparison, we also plot the
opacities measured at different redshifts in the spectra of Q 0302-003
(Heap et al. 2000), PKS 1935-692 (Anderson et al. 1999, reported
values come from Smette et al. 2002 who perform an optimal reduction
of the whole data set), and HE 2347-4342 (Smette et al. 2002). It is
clear from this figure that Model 5 provides a much better match to
the observational results.  This model also predicts full He {\small
II} reionisation by $z\simeq3.4$, a result consistent with the recent
analysis conducted by Theuns et al. (2002) who show clear evidence for
a sudden decrease in the effective optical depth in H {\small I} at
the same redshift due to the temperature increase associated with He
{\small II} reionisation. In the analysis presented in this paper, we
will adopt Model 5 as our most promising quasar model, based on its
predictions for the evolution of the He {\small II} opacity. In \S 7,
however, we summarise the related degeneracy associated with
successful quasar models and discuss the resulting implications in the
context of our study.

Given our QSO model, the {\it emergent} emissivity at each redshift
is calculated by summing up the B-band emissivity (in ergs
s$^{-1}$ Hz$^{-1}$ cm$^{-3}$) contributions from each source
and then using the universal SED to derive the following
expression for $\lambda<1050$ \AA:
\begin{equation}
\epsilon_{\rm QSO}(\nu,z)\simeq0.423 \ \epsilon_{\rm QSO}(\nu_B,z)\biggl
(\frac{\nu}{\nu_{1050}}\biggl)^{-1.9},
\end{equation}
where the numerical pre-factor accounts for the spectral mapping from
the blue frequency, $\nu_B$, to the frequency evaluated at
$\lambda=1050$ \AA, $\nu_{1050}$. In the following section, we
describe our prescription for adding a component from galaxies and define
an expression for its emissivity.

\section{GALAXIES}

Galaxies will represent the second class of sources which we allow to
contribute to the \ H {\small I} ionising background.  We assign a
spectral profile of the form $f(\nu)\propto \nu^{-\alpha_{\rm GAL}}$ for
$\lambda<912$\AA \ to these sources.  In this paper we assume
$\alpha_{\rm GAL}=5$, which is widely regarded as a realistic value based
on spectral studies of stellar atmospheres.  The much steeper spectral
slope relative to quasars means that galaxies produce an
insignificant number of He {\small II} ionising photons for any
reasonable luminosity model. This result justifies our approach of
relying on an He {\small II} reionisation simulation which contains
only quasars.  We emphasise that our conclusions are insensitive to
the value adopted for $\alpha_{\rm GAL}$ as long as galaxies contribute
negligibly to He {\small II} reionisation.

The next step is to adopt a functional form describing the redshift
evolution of the emissivity of this component.  Here we make the
assumption that the emissivity responsible for the ionising UV
background from this component is directly proportional to the
inherent star formation rate (SFR) within the galaxies.  This is a
fair premise since the SFR is a direct tracer of the young OB stars
which are the dominant producers of H {\small I} ionising photons in
galaxies. This assumption allows us to utilise empirical measurements
of the SFR at various redshifts as a basis for modeling the redshift
evolution of the emissivity within the redshift range in question.
In the future, it will also be interesting to contrast this with an 
analysis of direct theoretical predictions of the SFR (e.g. Nagamine 
et al. 2000, Springel \& Hernquist 2002a).
For our present purposes
it is convenient to parameterise the comoving {\it emergent}
emissivity from galaxies using the following expression:
\begin{equation}
\epsilon_{\rm GAL,c}(\nu,z)=\epsilon_{\rm GAL,c}(\nu_{\hh},3)\ f(z)\
\biggl(\frac{\nu}{\nu_{\hh}}\biggl)^{-\alpha_{\rm GAL}},
\end{equation}
where $\epsilon_{\rm GAL,c}(\nu_{\hh},3)$ represents the comoving
emissivity from galaxies at the hydrogen ionisation frequency
$\nu_{\hh}$ at $z=3$ and $f(z)$ is a dimensionless redshift dependent
function which is normalised to unity at $z=3$.  The function $f(z)$
can then serve to characterise the evolution of a particular galactic
model while the value of $\epsilon_{\rm GAL,c}(\nu_{\hh},3)$ provides
an overall normalisation.

In Figure 2, we present extinction-corrected data points for the
comoving SFR in a flat $\Lambda$ cosmology with parameters
($\Omega_m$, $\Omega_{\Lambda}$)=(0.37, 0.63) and $h=0.7$, as
provided by Nagamine, Cen, \& Ostriker (2000).  While not identical
to the cosmology we employ, that of Nagamine et al. is very similar to
ours and their summary of the observations is thus 
adequate for our present purposes.  As is
apparent, the sparse and loosely constrained observations beyond a
redshift $z\sim 2$ hardly warrant any attempt to fit the data as a means to
derive $f(z)$.  Rather, we chose the simple form
$f(z)=10^{m(z-3)}$ and consider two values for the slope, m, which
represent the subject of recent debates as to whether the SFR
continues to slightly rise or fall off beyond $z\sim3$ (see, e.g.
Madau et al. 1996, Steidel et al. 1999).  These models
are indicated by the bold short-dashed and long-dashed lines in
Figure 2, which we refer to as our galactic Model 1 (M1: m=-0.260) and 
galactic Model 2 (M2: m=0.135) respectively.  Our choices for the
slopes in these two cases are somewhat arbitrary, but they roughly bracket
the observations and will, therefore, enable
us to gauge the impact of a rising or falling SFR at $z > 3$ in 
the context of the analysis conducted in this paper. In both of these
models, we set
$\epsilon_{\rm GAL,c}(\nu_{\hh},3)=3.6\times10^{-49}$ ergs s$^{-1}$
Hz$^{-1}$ cm$^{-3}$ which is roughly $9\%$ smaller
than the corresponding quasar contribution. 
We discuss the motivation behind this choice in
\S 6 where we present results for the photoionisation rates.

\section{RECOMBINATION RADIATION FROM THE IGM}

Recombination radiation from the ionised IGM can also provide a
significant contribution to the ionising background.  In particular,
Haardt \& Madau (1996) have shown that this component can provide 
a fair fraction of the ionising
photons at redshifts near $z\sim3$. For helium, only
recombinations to the ground level of He {\small II} are able to
contribute to the diffuse He {\small II} ionising background. The helium
reionisation simulation used in this paper includes an approximate
treatment to account for this component which is directly incorporated
into the radiative transfer calculations (see \S 3.1.3 of Sokasian et
al. 2001).

For hydrogen, the following processes contribute to
the diffuse ionising background: (1) recombinations to the ground
state of H {\small I}, (2) He {\small II} Ly$\alpha$ emission
($2^2P\longrightarrow1^2S$) at 40.8 eV, (3) He {\small II} two-photon
($2^2S\longrightarrow1^2S$) continuum emission, (4) and He {\small II}
Balmer continuum emission at $\geq13.6$ eV. We ignore the 
contribution of He {\small I} and He {\small} II recombination
to the hydrogen ionising background. In the case of the former, this
is motivated by the relative smallness of the $n_{\he}/n_{\heh}$ and
$n_{\he}/n_{\hh}$ ratios encountered in typical intergalactic gas
which has been photoionised by galaxies and quasars. For the latter,
the exclusion is motivated by the fact that the relative cross section
for absorption of photons with energies $\geq$54.4 eV is much larger for He
{\small II} than H {\small I}.  Coupled with the fact that the typical
ratio for $n_{\hehh}/n_{\hh}$ encountered in the photoionised IGM is
large, we find that within the context of our analysis it is safe to
make the approximation that all the photons released from 
recombinations to the ground state of He {\small II} are absorbed by 
nearby He {\small II} ions before they have a chance to ionise H 
{\small I} atoms.

For the purposes of this paper, we find that it is sufficient to use
global number densities representative of the IGM to approximate the
corresponding emissivity from the relevant processes.  In order to
properly account for the inhomogeneity of the IGM, we compute global
volume-averaged clumping factors from radiative transfer grid at each
redshift, $C_f(z)$, which we then incorporate into the relevant
expressions.  We discuss the details associated with these
calculations in the following sections.

\subsection{Radiative Recombinations}
Given the electron number density, $n_e$, and the ion number density
$n_i$, the emissivity from direct recombinations to the $n^2L$ level
for hydrogen-like atoms with atomic number $Z$ and ionisation
threshold frequency $\nu_{th}$ from a photoionised gas that is in
local thermodynamic equilibrium at temperature, T, can be computed
using the Milne relation (Osterbrock 1989) which yields
\begin{equation}
\epsilon_{fb}(\nu)=\frac{4\pi}{c^2}\biggl(\frac{h^2}{2\pi
m_ekT}\biggl)^{3/2}n_en_i\frac{2n^3}{Z^3}h\nu^3\sigma_{\hh}(\nu/\nu_{th})e^{-h(\nu-\nu_{th})/kT},
\end{equation}
where $\sigma_{\hh}(\nu/\nu_{th})$ is the frequency dependent hydrogen
photoionisation cross section, $h$ and $k$ are the Plank and
Boltzmann constants, respectively, $m_e$ is the electron mass, and
$c$ is the speed of light.  Following the reasoning in Sokasian et
al. (2001) we artificially set all temperatures for the ionised
gas to $T=2.0\times10^4 K$ as a correction to the SPH temperatures
which exclude the extra heating introduced by radiative transfer
effects (see Abel \& Haehnelt 1999). As a further approximation, we set
$\sigma_{\hh}(\nu/\nu_{th})=\sigma_{\hh,o}(\nu/\nu_{th})^{-3}$ where
$\sigma_{\hh,o}=6.30\times10^{-18}$ cm$^{2}$ is the photoionisation
cross section at the Lyman limit for H {\small I} (Osterbrock 1989).

To compute realistic values for the recombination emissivity, we
require information regarding the clumping statistics associated with
the IGM.  In the context of our calculation, this will reduce to a
global volume-averaged clumping factor for each redshift, $C_f(z)$,
which we can then use as a multiplicative prefactor in the above
expression. In order to obtain reliable values for the clumping factor
of the IGM from our radiative transfer grid, it is necessary to discount 
cells
harbouring collapsed objects with cold gas that are not part of
the IGM but which can significantly alter the clumping statistics. We
achieve this by considering only cells below a specific overdensity
cut-off. The choice for the cut-off is somewhat arbitrary since the
distinction between the IGM and collapsed objects is
blurred.  In this paper, we employ scatter plots showing the density 
vs. temperature for the SPH particles (see, e.g., Dav\'e et al. 1999 
or Springel \& Hernquist 2002b) to estimate this cut-off. In particular,
these plots show a bifurcation between the reservoir of underdense cool
gas (i.e. the IGM) and shock-heated gas of moderate density or cold high 
density gas in collapsed objects which occurs somewhere between an
overdensity of $10-50$, independent of redshift. In our analysis, we
adopt a median value of $30$ for our overdensity cut-off at
all redshifts. This value results in an exclusion of only $0.14\%$ of
the volume and a global volume averaged clumping factor of $3.88$ at
$z=3$.  Adopting an overdensity cut-off of 10 (50) would lead to a 
$-32\%$ ($+27\%$) change in the clumping factor at the same redshift.
Although only $0.14\%$ of the volume is excluded at $z=3$ with the 
adopted value for the cut-off, the corresponding mass harboured in these
excluded cells corresponds to roughly $11.5\%$ of the total mass in the
simulation volume. For consistency, we exclude the corresponding 
{\it collapsed-mass} at each redshift when deriving $n_e$ and $n_i$.

We note that our estimates for the clumping factor are considerably
smaller than those of Springel \& Hernquist (2002a; as summarised by
e.g. their Figure 16), because here we exclude high density regions
from our estimate of $C_f(z)$.  This is appropriate because in this
paper we are interested in the volume averaged properties of the IGM
at a redshift when the Universe was essentially optically thin to
hydrogen ionising photons, while Springel \& Hernquist (2002a) were
concerned with the situation leading to hydrogen reionisation when
most of these photons would have been absorbed in the vicinity of the
sources which produced them.  In any event, as indicated above, a
modest increase in the overdensity threshold would have a similarly
modest impact on the clumping factor and the inferred emissivity from
radiative recombination.

For simplicity, the latter densities are approximated by assuming
complete ionisation ($n_i/n_{\rm tot}=1$) within the photoionised
regions.  This approximation is certainly justified in comparison to
other uncertainties present in our analysis. The
relevant densities are thus given by,
\begin{equation}
\begin{align}
n_e(z) & = 2n_{He,{\rm tot}}I_{\hehh}+n_{H,{\rm tot}} I_{\hhh}(z) \\ 
n_{\hh}(z) & = n_{H,{\rm tot}} I_{\hhh}(z) \\ 
n_{\hehh}(z) & = n_{He,{\rm tot}} I_{\hehh}(z)
\end{align}
\end{equation}
where $n_{i,{\rm tot}}$ represents the total number density of species $i$
averaged over the entire simulation box (excluding the collapsed-mass
as defined above), and $I_{i}(z)$ is the fraction of the volume which
has become ionised in species $i$ by redshift $z$.  In the case of He
{\small III}, $I_{\hehh}(z)$ is extracted directly from the helium
simulation results. For H {\small II}, we approximate $I_{\hhh}(z)$ as
the ratio of the cumulative number of H {\small I} ionising photons
that were released by quasars and galaxies by redshift $z$ to the
total number of hydrogen atoms present in the IGM (which again
excludes the collapsed-mass fraction). Obviously, we restrict the
maximum value of $I_{\hhh}(z)$ to unity. It is also important to point
out that we limit contributions to the H {\small I} ionising pool
only to photons with frequencies in the range
$\nu_{\hh}<\nu<\nu_{\heh}$ under the earlier premise that photons with
frequencies $\nu > \nu_{\heh}$ are immediately absorbed in the IGM by
the He {\small II} ions.

\subsection{Ly$\alpha$ Emission}
Recombinations into He {\small II} which end up populating the $2^2P$
level are converted to Ly$\alpha$ 304 \AA \ photons which are capable
of ionising H {\small I}. These photons resonantly scatter,
and therefore diffuse only slowly away from their point of
origin before they are absorbed.  As a result, the immediate fate of a
He {\small II} Ly$\alpha$ photon mainly depends upon the competition
between H {\small I} continuum absorption and the local opacity at the He
{\small II} Ly$\alpha$ frequency $\nu_{\alpha}$ (the effect of dust
on destroying Ly$\alpha$ radiation is negligible [Haardt \& Madau
1996]).  For the purposes of our analysis, it is sufficient to assume
that all Ly$\alpha$ photons eventually contribute to the H {\small I}
ionising background either by scattering enough times to encounter an H
{\small I} atom in the cloud of origin or eventually redshifting below
the line frequency and escaping into the IGM.  Which process comes
first depends on the velocity gradient versus the absorption
coefficient of H {\small I} at the He {\small II} Ly$\alpha$
frequency.  In either case, assuming all the energy is released
exactly at the line frequency, the emissivity associated with
Ly$\alpha$ line radiation can be written as,
\begin{equation}
\epsilon_{{\rm Ly}\alpha}(\nu)=h\nu\delta(\nu-\nu{_\alpha})n_{2^2P}A_{2^2P,1^2S},
\end{equation}
where $A_{2^2P,1^2S}$ is the transition probability for
$2^2P\longrightarrow1^2S$ transitions in He {\small II} and $n_{2^2P}$
is number density of He {\small II} ions in the $2^2P$ state.  In our
case, the transition probability cancels out in the above
expression since we derive $n_{2^2P}$ by assuming the equilibrium condition,
\begin{equation}
0.75\alpha_{B}n_en_{\hehh}=n_{2^2P}A_{2^2P,1^2S},
\end{equation}
where the use of the Case B recombination coefficient, $\alpha_{B}$,
implicitly assumes that all $n>2$ recombinations will
eventually cascade down and populate the $n=2$ level.  For this paper,
we use the fitting formula provided by Hui \& Gnedin (1997) and arrive
at $\alpha_{B}=9.089\times10^{-13}$ cm$^3$ s$^{-1}$ for
$T=2.0\times10^4K$.  The factor 0.75 represents the fraction of
recombinations to the excited states that will eventually populate the
$2^2P$ state based on the degeneracy of available states in the $P$
level (the remainder end up in the $2^2S$ state).  This assumes that
the rate of transitions between the $S$ and $P$ states is small, which
is valid at the typical densities associated with the IGM. 

\subsection{Two-Photon Continuum}
The radiative decay $2^2S\longrightarrow1^2S$ in He {\small II} is
almost entirely due to two-photon emission and is also capable of
contributing to the diffuse H {\small I} ionising background. The
emissivity for this process can be expressed as,
\begin{equation}
\epsilon_{2ph}(\nu)=\frac{h\nu}{\nu_{\alpha}}A_{2^2S,1^2S}(\nu/\nu_{\alpha})n_{2^2S},
\end{equation}
where $A_{2^2S,1^2S}(\nu/\nu{_{\alpha}})$ is the frequency dependent
transition probability, which again cancels out via the assumption of
the equilibrium condition,
\begin{equation}
0.25\alpha_{B}n_en_{\heh}h=n_{2^2S}A_{2^2S,1^2S},
\end{equation}
adopted under the premise mentioned in the preceding section.

\section{COMPUTING THE IONISING BACKGROUND}
Given the emissivities of each component, we can proceed to the
calculation of the resulting ionising background intensity and the
corresponding photoionisation rate.  The cosmological radiative transfer
equation for diffuse radiation can be expressed as (see, e.g., Peebles
1993),
\begin{equation}
\biggl(\frac{\partial}{\partial
t}-\nu\frac{\dot{a}}{a}\frac{\partial}{\partial
\nu}\biggl)J=-3\frac{\dot{a}}{a}J-c\kappa J+\frac{c}{4\pi}\epsilon,
\end{equation}
where $a$ is the scale factor, $\kappa$ is the continuum absorption
coefficient per unit length along the line of sight, and $\epsilon$ is
the proper space-averaged volume emissivity which in our case can be
expressed as the sum
$\epsilon(\nu,z)=\epsilon_{\rm QSO}(\nu,z)+\epsilon_{\rm GAL}(\nu,z)+\epsilon_{\rm IGM}(\nu,z)$
where
$\epsilon_{\rm IGM}(\nu,z)=\epsilon_{fb}(\nu,z)+\epsilon_{{\rm Ly}\alpha}(\nu,z)+\epsilon_{2ph}(\nu,z)$.  The
mean specific intensity of the background, $J$, at the observed
frequency $\nu_o$, as seen by an observer at redshift $z_o$ is then,
\begin{equation}	
J(\nu_o,z_o)=\frac{1}{4\pi}\int^{\infty}_{z_o}\frac{dl}{dz}
\frac{(1+z_o)^3}{(1+z)^3}\epsilon(\nu,z)e^{-\tau_{\rm eff}} \ dz,
\end{equation}	
where $\nu=\nu_o(1+z)/(1+z_o)$ and $dl/dz$ is the proper line element
in our $\Lambda$CDM cosmology,
\begin{equation}
\frac{dl}{dz}=\frac{c}{H_o
(1+z)}[0.3(1+z)^3+0.7]^{-1/2},
\end{equation}
where c is the speed of light and $H_o$ is the present day Hubble
parameter.  The remaining exponential term accounts for
absorption occurring through $dz$ due to discrete absorption systems
which is parameterised by a mean optical depth $\tau_{\rm eff}$ which, for
a Poisson-distribution of clouds, can be expressed as:
\begin{equation}
\tau_{\rm eff}(\nu_o,z_o,z)=\int_{z_o}^z
dz^{\prime}\int_0^{\infty}\frac{\partial^2N_{\rm col}}{\partial
N_{\rm col,\hh} \partial z^{\prime}} (1-e^{-\tau})\ dN_{\rm col,\hh}
\end{equation} 
(Paresce, McKee, \& Bowyer 1980), where $\partial^2N_{\rm col}/\partial
N_{\rm col,\hh} \partial z^{\prime}$ is the redshift and column density
($N_{\rm col}$) distribution of absorbers along a line of sight, and
$\tau$ is the Lyman continuum optical depth through an individual
cloud.  The usual form for the redshift and column density
distribution of absorber lines is given as:
\begin{equation}
\frac{\partial^2N_{\rm col}}{\partial N_{\rm col,\hh} \partial z}=A_o \
N_{\rm col,\hh}^{-1.5}(1+z)^{\gamma}.
\end{equation}
with $A_o$ and $\gamma$ acting as fitting parameters.  For our
analysis here, we choose to fit the above function exactly as in
Madau, Haardt, \& Rees (1999) where a single redshift exponent,
$\gamma=2$, is assumed for the entire range in column densities with
a normalisation value of $A_o=4.0\times 10^7$. At $z=3$, the adopted
values produce roughly the same number of Lyman limit systems and
lines above $N_{\rm col,\hh}=10^{13.77}$ cm$^{-2}$ as observed by
Stengler-Larrea et al. (1995) and estimated by Kim et al. (1997),
respectively.  With a single power law describing the distribution of
absorbers along a line of sight, the effective optical depth can now
be expressed as an analytic function of redshift and frequency
(i.e. equation [6] in Madau et al. 1999),
\begin{equation}
\tau_{\rm eff}(\nu_o,z_o,z)=\frac{4}{3}\sqrt{\pi \sigma_{\hh,o}} N_o
(\nu_o/\nu_{\hh})^{-1.5} (1+z_o)^{1.5}[(1+z)^{1.5}-(1+z_o)^{1.5}].
\end{equation}  
It is important to note that the expression above for the mean opacity
does not explicitly include the contribution of helium to the
attenuation.  While the opacity due to He {\small I} ionisations at 504
\AA \ is negligible in the case of a background composed in part by
hard sources like quasars (Haardt \& Madau 1996), He {\small II}
absorption may still contribute a measurable level of opacity for the
population of photons with frequencies $\nu > \nu_{\heh}$ (produced
almost exclusively by quasars).  However, as we shall show below, any
additional attenuation due to helium absorption has a very small
effect on the H {\small I} photoionisation rate and does not affect
the results of this analysis; we therefore omit this component.

Given the background intensity $J(\nu,z)$, the global
photoionisation rate can then be calculated according to:
\begin{equation}
\Gamma_{\hh}(z_s)=\int_{\nu_{\hh}}^{\infty}\frac{4\pi
J(\nu,z)}{h\nu}\sigma_{\hh,o}(\nu/\nu_{\hh})^{-3} \ d\nu
\end{equation}
where we have a adopted a frequency dependence of
$(\nu/\nu_{\hh})^{-3}$ for the photoionisation cross section.  Here we
note that for the quasar model employed in this paper, an integration
up to only $\nu_{\heh}$ would have decreased the photoionisation rate
by roughly $0.2\%$ at z=3.0. This thus represents the maximum
decrement any additional attenuation from helium absorption may cause
and is small enough for us to continue with its omission.

\section{RESULTS AND DISCUSSION}

In Figure 3, we plot the resulting background intensities at
$\lambda=912$ \AA \ for three cases: 1) quasars alone, 2) quasars
combined with an M1 galactic component, and 3) quasars combined with an
M2 galactic component.  In both galactic models, we set
$\epsilon_{\rm GAL,c}(\nu_{\hh},3)=3.4\times10^{-49}$ ergs s$^{-1}$
Hz$^{-1}$ cm$^{-3}$, which is roughly equal to the corresponding
contribution from the quasars at the same redshift. The associated
recombination component from the ionised IGM is also included in all
three cases. The values plotted for the quasar contribution reflect an
average over 20 unique realisations associated with the quasar section
algorithm.  The shaded region refers to the Lyman limit background
estimated from the proximity effect (Giallongo et al. 1996; Cooke et
al. 1997; Scott et al. 2000). Here we see that all three cases produce
total intensities that appear to be consistent with measurements,
although it is obvious that observational uncertainties
are quite large.  Nevertheless, it is arguable that the
larger intensities associated with the inclusion of the galactic
component M2 appear to offer the best agreement with the measurements
at $z\gtrsim3.8$.

The resulting H {\small I} photoionisation rate for each of the three
cases is plotted in Figure 4. Hatched regions represent the range of
values corresponding to the 90\% confidence level from the 20 
realisations of the quasar sources. The two data points are
measurements of $\Gamma_{\hh}$ over the indicated redshift ranges 
(horizontal lines) given by Rauch et al. (1997), who measured the 
H {\small I} Ly$\alpha$ forest absorption from intervening gas in
seven high resolution QSO spectra obtained with the
Keck telescope. The values of these data points have been adjusted to the
cosmology used in this paper. Here we can clearly see that quasars
alone, although successful in matching $z\simeq3$ measurements, fail
to produce the observed rates in the range $3.5<z<4.5$ by a large
factor. The inclusion of the M1 galactic component whose contribution
tapers off with redshift, does not appear to do any better and is,
in fact, incompatible with the $2.5<z<3.5$ measurement as well.  Only
by including an M2 galactic component do we obtain a reasonable
match to the observations.

These conclusions have potential implications for our understanding of
recent observations indicating that hydrogen in the Universe was
reionised by redshift $z\sim 6$ (e.g. Becker et al. 2001).  In
particular, from Figure 3c, we see that the background intensity at
the hydrogen Lyman limit predicted by our model becomes increasingly
dominated by stars for $z > 4$.  This suggests that unless there are
other sources of ionising radiation present in the real Universe that
have been neglected in our analysis, such as mini-quasars or weak
active galactic nuclei, hydrogen reionisation at $z\sim 6$ must have
been driven mainly by stellar radiation.

The fact that our choice for $\epsilon_{\rm GAL,c}(\nu_{\hh},3)$ results
in a reasonable match to the photoionisation measurements in the latter
case is not coincidental as it was specifically chosen for this
purpose. To investigate whether this value is reasonable in the context
of corresponding measurements of the emissivity of galaxies at larger
wavelengths ($\lambda\simeq1500$ \AA) we will need to first adopt
a realistic ratio between the flux densities at 1500 \AA \ and 900 \AA,
$f[1500]/f[900]$. After analysing a composite spectrum of 29 LBGs at
$z\sim3.4$, Steidel et al. (2001) derived an observed ratio of
$f[1500]/f[900]\simeq17.7$. However, it should be emphasised that the
LBGs comprising their composite spectrum were drawn from the bluest
quartile of intrinsic far-UV colours and may be exhibiting larger than
average 900 \AA \ continuum emission.  In fact, more recently,
Giallongo et al. (2002) examined the spectra of two galaxies at
$z=2.96$ and $z=3.32$ which exhibited little or no flux at
$\lambda=900$ \AA \ and were not included in the latter
subsample. They derived a lower limit of $f[1500]/f[900]>71$ which is
roughly 4 times the value derived by Steidel et al. (2001). 

In Figure 5 we plot the resulting comoving emissivity at
$\lambda=1500$ \AA \ from our galactic model M2 as a function of
redshift based on the two values of the observed flux ratio from
Steidel et al. (2001) and Giallongo et al. (2002). Also plotted are
the relevant observations of the UV emissivity at $\lambda=1500$ \AA \
from Steidel et al. (1999), Pascarelle et al. (1998), Madau et
al. (1996), and Madau et al. (1998) adjusted for a flat $\Lambda$CDM
cosmology similar to ours. With the exception of the Madau et al. data
point at $z=4$, it appears as if the adopted galactic model M2 offers
good agreement with the observations if the typical value for
$f[1500]/f[900]$ lies somewhere between the values advocated by
Steidel et al. (2001) and Giallongo et al. (2002).  This behaviour is
consistent with the recent theoretical predictions of Springel \&
Hernquist (2002a), who find that the SFR in high resolution
simulations which include hydrodynamics and a multi-phase model for
star forming gas (Springel \& Hernquist 2002c)
rises from $z=0$ out to $z\approx 5.4$ before
declining at even higher redshifts.

Our value of $\epsilon_{\rm GAL,c}(\nu_{\hh},3)$ also appears to be
consistent with the recent suggestion that LBGs emit a
comparable number of ionising photons to QSOs at $z\sim3$ (Steidel
et al. 2001), although we remind the reader that the adopted QSO model
(Model 5 from SAH) in this paper has less emission than the
``standard'' QSO model (Model 1 from SAH), as is required
in order to match the He {\small II} opacity measurements (see Table
1, Figure 1).  More specifically, in the case which combines the
adopted QSO model and galactic Model M2 we find
$J_{\rm GAL}(\nu_{\hh})/J_{\rm QSO}(\nu_{\hh})\simeq0.77$ at $z\simeq3$
whereas the same galactic component combined with the standard QSO
model would produce $J_{\rm GAL}(\nu_{\hh})/J_{\rm QSO}(\nu_{\hh})\simeq0.47$
at the same redshift.  It is interesting to point out that the standard
QSO model not only proves to be unsuccessful in matching the helium
opacity measurements but would also correspond to an unsuccessful
model in the context of the hydrogen photoionisation rate measurements by
Rauch et al. (1997).  In particular, without any additional
contribution from a galactic component, the resulting photoionisation
rate from the standard QSO model would overpredict the $2.5<z<3.5$
measurement by $\simeq43\%$ while underpredicting the $3.5<z<4.5$
measurement by $\simeq59\%$. Furthermore, any attempt to include a
reasonable contribution from a galactic component, such as in Model M2,
would inevitably exacerbate the disagreement at $z\simeq3$.  It therefore
appears as if the standard QSO model as defined by the parameters in
Table 1 for Model 1 in conjunction with the Pei (1995) $B$-band fit to
the luminosity function presented by Boyle et al. (1988) overpredicts
the ionising emissivity in both hydrogen and helium at
$z\simeq3$.  This conclusion is consistent with the preliminary
analysis of the density of faint QSOs carried out by Steidel et al. (1999)
in which their LBG survey indicated that the standard extrapolated QSO
luminosity function may slightly overpredict the QSO contribution to
$J(\nu_{\hh})$ at $z\sim3$.

If we are to accept the form and fit to the QSO luminosity function,
then the incompatibility between the standard QSO model and
observations suggests
that either the extrapolation to the faint end or the adopted SED or a
combination of both is misrepresented.  The extrapolation to faint
galaxies is parameterised by the minimum luminosity allowed for the
quasars, $L_{min}$.  The widely adopted procedure for estimating
$L_{min}$ is based on the idea that present day Seyfert galaxies are
counterparts to high redshift
quasars and that the faintest luminosities associated
with Type I Seyferts should be representative of the minimum
luminosities of quasars once a luminosity evolution model has been
factored in.  A logical choice for the evolution model, which we
subsequently adopted in this paper, is based on the observed evolution
of the break luminosity in the LF of quasars.  However, as the failure
of the standard model suggests, adopting the faintest values for
$L_{min,0}$ from Seyfert galaxies appears to produce too many faint
QSOs and is partly responsible for the apparent overproduction of
ionising photons in both H {\small I} and He {\small II}. The implication
then could be that the minimum luminosity evolves separately from the
apparent evolution in the break luminosity.  A possible motivation for
such a scenario could be provided by theories of hierarchical structure
formation which are more efficient in inhibiting the formation of the
smallest supermassive black holes and/or the accretion rate onto
them.  Delving into the theory related to such speculations is beyond
the scope of this paper; instead we refer the reader the work of Haiman \&
Menou (2000) and Kauffmann \& Haehnelt (2000) which provide a good review
of the subject.

With respect to the form for the SED, it is important to point out
that the EUV spectral indices observed in quasar spectra exhibit a
significant amount of scatter (see, e.g, Zheng et al. 1998, Telfer
et al. 2002), which makes it difficult to define a
universal index representative of all quasars. This dispersion is
apparent in Figure 6 where we show the combined distribution of EUV
indices for radio-quiet and radio-loud QSOs from Telfer et al. (2002)
(data kindly provided by G. Kriss). The indices in this distribution
are defined by $f(\nu)\propto\nu^{-\alpha}$ between 500 and 1200 \AA,
which is comparable to our definition for $\alpha_{QSO}$. It is clear
from this plot and the corresponding statistics that it is difficult
to represent this quantity by a single value appropriate for
all quasars.  Within this context,
our choice of $\alpha_{QSO}=1.9$ seems entirely plausible. We must
note, however, that the situation would be much more convoluted if
there existed some correlation between the spectral index and the
intrinsic luminosity of the source, as suggested by the strong
luminosity evolution in the X-ray and optical wave bands of quasars
(see, e.g., Boyle 1994).  Recent investigations into this
matter (Yuan et al 1998; Brinkmann et al. 1997) have revealed that the
observed $\alpha$-luminosity correlation can be attributed to the
dispersion in the observational data points and is thus not an
underlying physical property of the sources.

\subsection{Other QSO Models} 
We have shown that a QSO model with a larger value for
the minimum luminosity at $z=0$ and a slightly steeper tail-end slope
than the standard model, can simultaneously provide a fair match to
the observed helium opacity and hydrogen photoionisation rate
measurements when one includes a realistic soft contribution from
galaxies.  It is important to point out, however, that the specific
examples chosen in this paper represent a degenerate class of models
which can also be combined to produce similar results.  More
specifically, in the case of quasars, the parameters $L_{\rm min,0}$ and
$\alpha_{QSO}$ can both be adjusted to deliver the required ionising
emissivity in helium while changing the resultant ionising emissivity
in hydrogen. This would therefore create a large set of different
galactic models which would also be fairly successful at reproducing
the observations. In Figure 7, we demonstrate this degeneracy by
plotting the absolute value of the difference between the logarithms of
the computed hydrogen photoionisation rate ($\Gamma_{\hh}$) and the
measured value obtained by Rauch et al. (1997) ($\Gamma_{\hh,m}$) as
a function of quasar parameters $L_{\rm min,0}$ and $\alpha_{\rm QSO}$ at
redshifts 3 and 4.  The labeled contours in each panel show the
percent difference from the corresponding measured value.  To tie in
the correlation with the helium observations, we also plot in each
panel the curve ({\it white-dashed line}) in the
$L_{\rm min,0}$-$\alpha_{\rm QSO}$ plane that produces the same number of He
{\small II} ionising photons as quasar Model 5 from SAH. The success
of the latter model in matching He {\small II} opacity measurements
and the strong correlation between the number of ionising photons
released and the resulting opacities, effectively restricts us to
consider models only in the vicinity of this curve. The plots make it
apparent that a pure quasar model based on the adopted LF, while
capable of producing agreement at $z=3$, fails by a large factor
at $z=4$ for any reasonable range in $L_{\rm min,0}$ and
$\alpha_{\rm QSO}$. On the other hand, as the bottom panels indicate, a
quasar model which is supplemented with an additional soft component,
which in this case is galactic Model M2, is able to significantly
reconcile this failure.  It is important to point out to the
reader that Figure 7 is based on relatively sparse data in both H
{\small I} and He {\small II} (7 and 4 quasar spectra, respectively)
and that more observations, especially at high redshift, are necessary
to place definitive constraints on the models.

\section{CONCLUSION}

In this paper, we have utilised observations in both H {\small I} and
He {\small II} to estimate the contributions to the UV background from
quasars and galaxies.  The fact that only quasars are capable of
producing radiation hard enough to ionise He {\small II} has allowed
us to select a particular quasar model based solely on the He {\small
II} opacity measurements, independent of the galactic model. Including
measurements of the photoionisation rate in H {\small I} between
$2.5\simlt z\simlt 4.5$ then enables us to study predictions based on two
hotly debated models for the galactic contribution.  

We find that a quasar model with less emission than the widely quoted
``standard'' model in conjunction with a galactic model with a
slightly rising SFR that contributes a comparable amount of H {\small
I} ionising radiation at $z\simeq3$, is necessary to achieve good
agreement with all relevant observations.  Such a composite model
makes it much easier to understand how the intergalactic medium can
remain highly ionised at redshifts $z>4$ where the contribution from
bright quasars falls off significantly.  It can also explain the
apparent progressive softening of the UV background at $z>3$ as
suggested by metal absorption line observations (Savaglio et al. 1997;
Songaila 1998).  The particular choice for the galactic component in
the above model is further bolstered by the fact that the galactic
model appears to match observations for the emissivity of LBGs between
$2.5\simlt z\simlt 4.5$ as it appears as if the adopted galactic model M2
offers good agreement with the observations if the typical value for
$f[1500]/f[900]$ lies somewhere between the values measured by Steidel
et al. (2001) and Giallongo et al. (2002).  Moreover, the rise in the
SFR beyond $z\sim 3$ is in accord with some observations and theoretical
predictions (for a discussion, see e.g. Springel \& Hernquist 2002a).

We have also shown that there exists a degenerate class of quasar models
which are equally successful at matching the He {\small II}
observations while producing a large dispersion in their H {\small I}
contributions.  However, the particular QSO model adopted in this paper
has the interesting property of being characterised by plausible values
for $L_{\rm min}$ and $\alpha_s$ while naturally requiring an extra
galactic component that seems to be consistent with observations of
both its amplitude and shape.

While the scenario presented in this paper appears promising, we must
emphasise that it is based on relatively sparse data.  Future
observations gathered with the Sloan Digital Sky Survey (SDSS) will
allow us to reduce the uncertainties associated with quasar models at
high redshifts.  Coupled with future measurements of the proximity
effect and the evolution of the intensity ratio of metal lines, it should
soon be possible to place even tighter constraints on the relative
contributions from quasars and galaxies to the UV background at 
$z\sim 2.5 - 5$.

\begin{acknowledgments}
We thank Kentaro Nagamine for providing us with data related to the
SFRs of LBGs in a convenient and useful form as well as providing
useful comments regarding the nature of galactic sources.  We also
thank Volker Springel for comments on the manuscript,
Martin Elvis for informative discussions concerning the specifics
of quasar SEDs in the context of our analysis, Kurt Adelberger for
useful discussions about the UV emissivity of LBGs, and George
Rybicki for discussions related to radiative processes.  A.S. thanks
Daniel Harvey for many constructive discussions related to this
study. This work was supported in part by NSF grants ACI96-19019,
AST-9803137, and PHY 9507695.
\end{acknowledgments}

\clearpage

\begin{table}[htb]
\begin{center}
\begin{tabular} {cccccc}
\multicolumn{6}{c}{\textbf{TABLE 1 }} \\
\multicolumn{6}{c}{Quasar Source Parameters } \\
\hline
\hline
Quasar Model & $T_{\rm life}$ [$10^7$ yrs] & $L_{\rm min,0}$ [$10^{9} L_{B,\odot}$]  &$M_{\rm min}$ [$10^{10} M_{\odot}$] & $\beta$ [radians] & $\alpha_{\rm QSO}$ \\
\hline
1 & 2.0 & 3.91 & 1.8 & $\pi$ & 1.8\\
5 & 2.0 & 23.5 & 1.8 & $\pi$ & 1.9\\
\hline
\end{tabular}
\end{center}
\end{table}

\clearpage

\begin{figure}[htb]
\figurenum{1}
\setlength{\unitlength}{1in}
\begin{picture}(6,6.5)
\put(-0.65,-1.7){\includegraphics{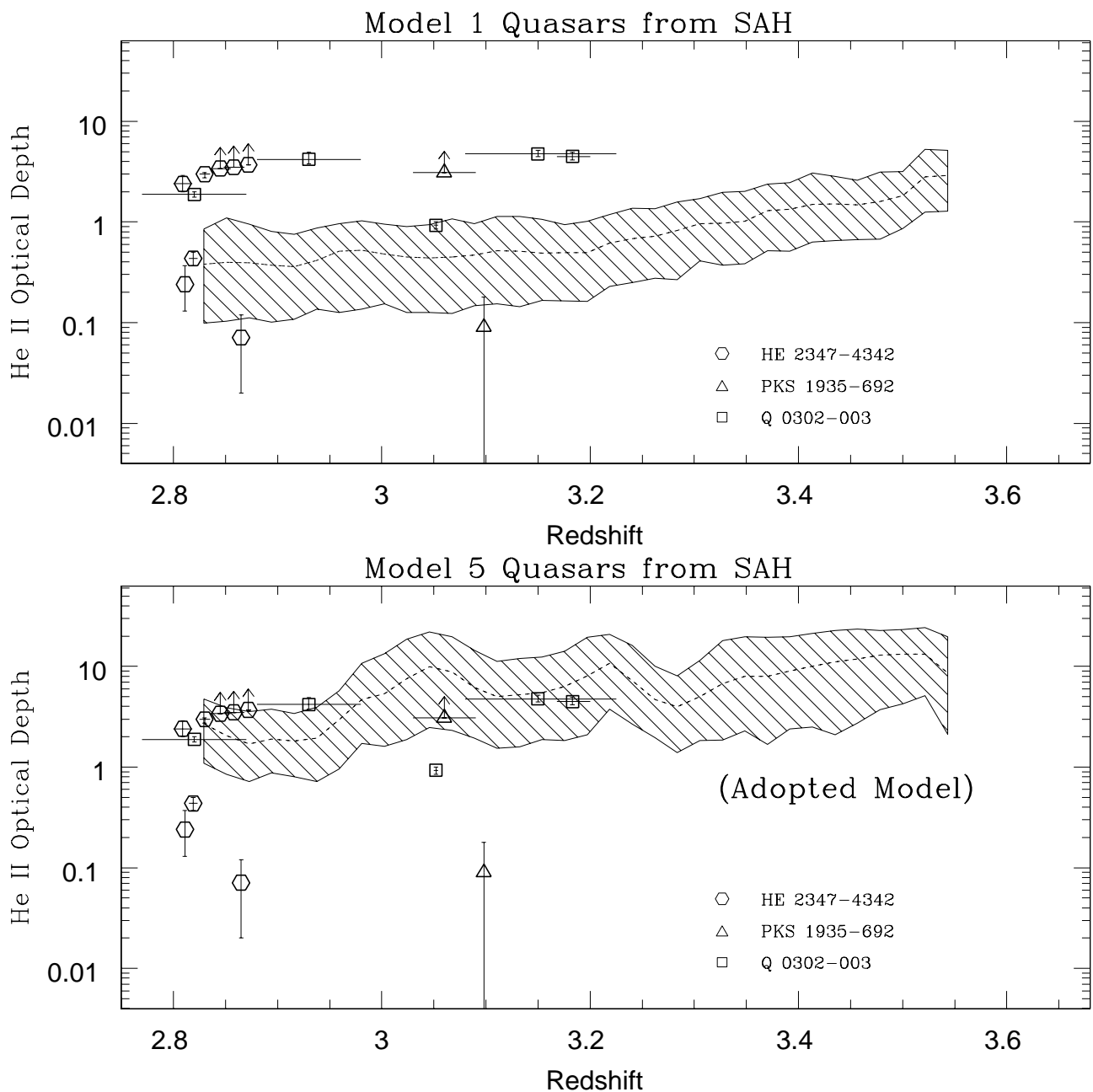}}
\end{picture}
\caption{Redshift evolution of the effective mean optical depth of He
{\small II} absorption in Models 1 and 5 from SAH. Hatched regions
represent the optical depth derived from the simulations at the 90\%
confidence level with the dashed lines indicating mean
values. Observational results from quasars HE 2347-4342, PKS 1935-692
and Q 0302-003 are plotted for comparison. We adopt Model 5 for our
analysis in this paper.}
\end{figure}

\begin{figure}[htb]
\figurenum{2}
\setlength{\unitlength}{1in}
\begin{picture}(6,6.5)
\put(-0.65,-1.7){\includegraphics{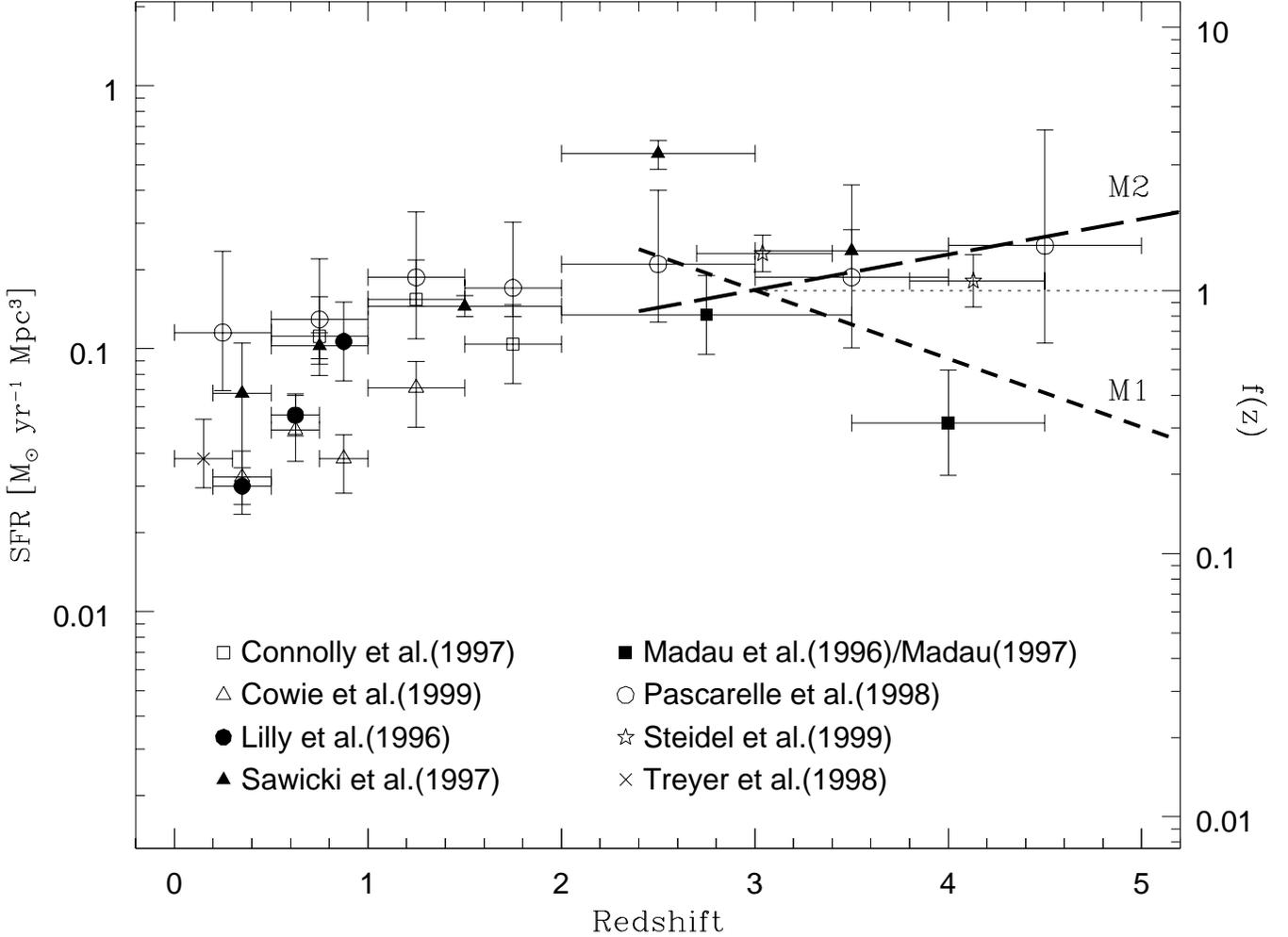}}
\end{picture}
\caption{Extinction-corrected data points for the comoving SFR in a
flat $\Lambda$ cosmology with parameters 
($\Omega_m$, $\Omega_{\Lambda}$)=(0.37, 0.63)
and $h=0.7$ as summarised by Nagamine, Cen, \& Ostriker (2000). Sources
for the raw data are listed above according to their corresponding
points. The bold short-dashed and long-dashed lines represent Models 1
(M1) and 2 (M2) for our comoving redshift evolution factor $f(z)$
(right axis), respectively. Note that $f(z)$ is normalised to unity at
$z=3$ for both models, in accord with our parameterisation of the
comoving galactic emissivity:
$\epsilon_{\rm GAL,c}(\nu,z)=\epsilon_{\rm GAL,c}(\nu_{\hh},3)\ f(z)\
(\nu/\nu_{\hh})^{-\alpha_{GAL}}.$}
\end{figure}

\begin{figure}[htb]
\figurenum{3}
\setlength{\unitlength}{1in}
\begin{picture}(6,6.5)
\put(0.81,-2){\includegraphics{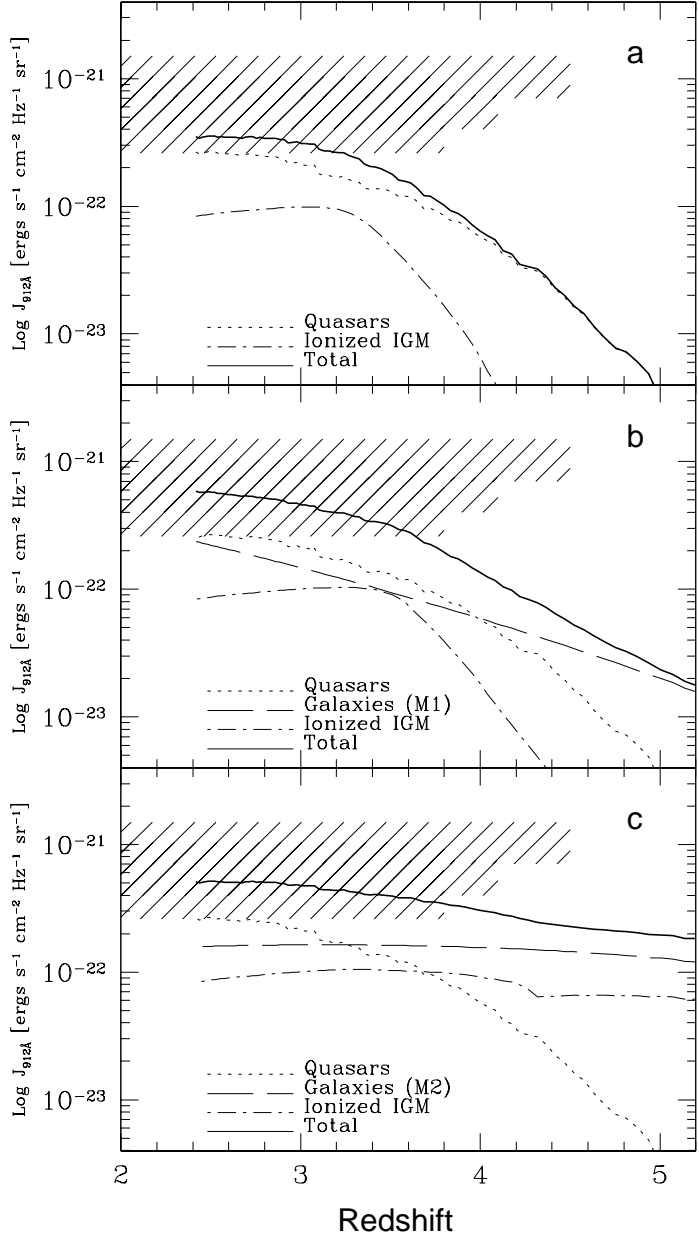}}
\end{picture}
\caption{Background intensity at $\lambda = 912$ \AA \ resulting from
(a) quasars only, (b) quasars and galactic Model M1, and (c) quasars
and galactic Model M2. Shown are the separate contributions from
quasars (dotted), galaxies (dashed), and the ionised IGM
(dashed-dotted). The shaded region refers to the Lyman limit
background estimated from the proximity effect (Giallongo et al. 1996;
Cooke et al. 1997; Scott et al. 2000)}
\end{figure}

\begin{figure}[htb]
\figurenum{4}
\setlength{\unitlength}{1in}
\begin{picture}(6,6.5)
\put(0.81,-2){\includegraphics{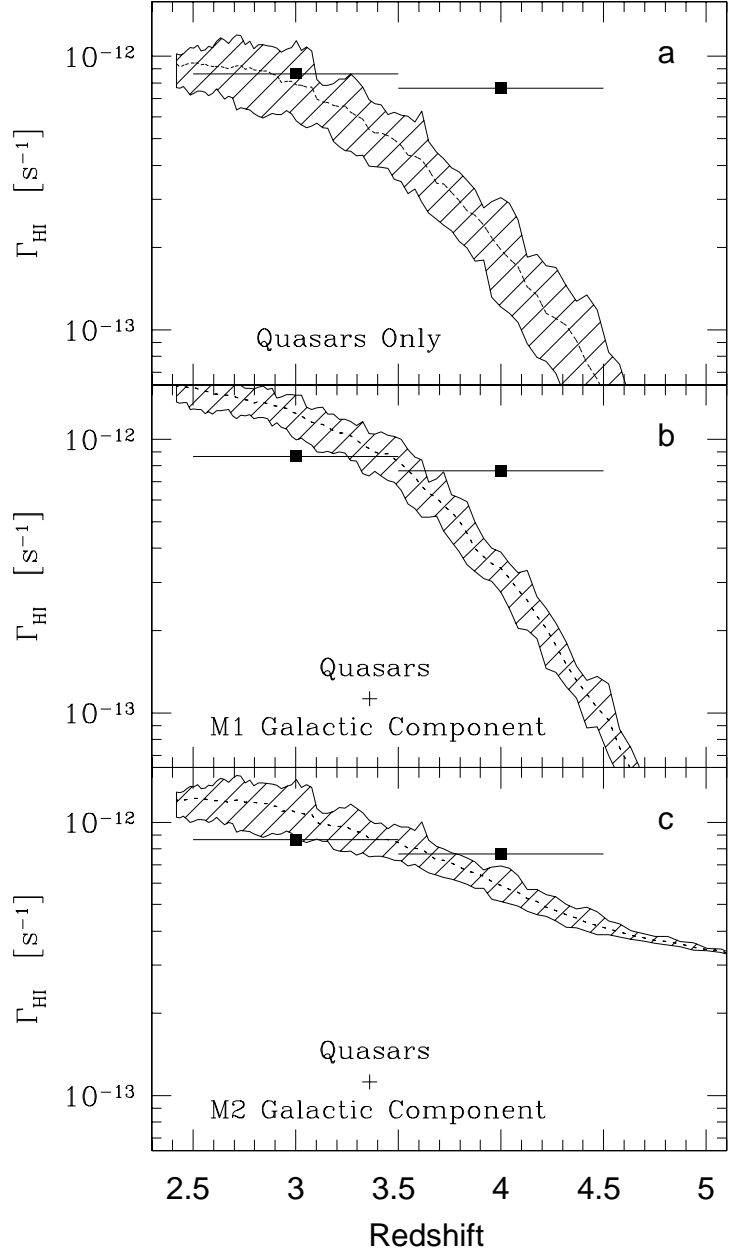}}
\end{picture}
\caption{Hydrogen photoionisation rate as a function of redshift
resulting from (a) quasars only, (b) quasars and galactic Model M1,
and (c) quasars and galactic Model M2.  Hatched regions represent the
range of values corresponding to the 90\% confidence level from repeating
the analysis with 30 unique random seeds associated with the quasar
selection algorithm.  The two data points represent measurements
of $\Gamma_{\hh}$ over the indicated redshift ranges (horizontal
lines) obtained by Rauch et al. (1997), after being adjusted for the
cosmology used in this paper (see text).}
\end{figure}

\begin{figure}[htb]
\figurenum{5}
\setlength{\unitlength}{1in}
\begin{picture}(6,7.0)
\put(-0.65,-1.7){\includegraphics{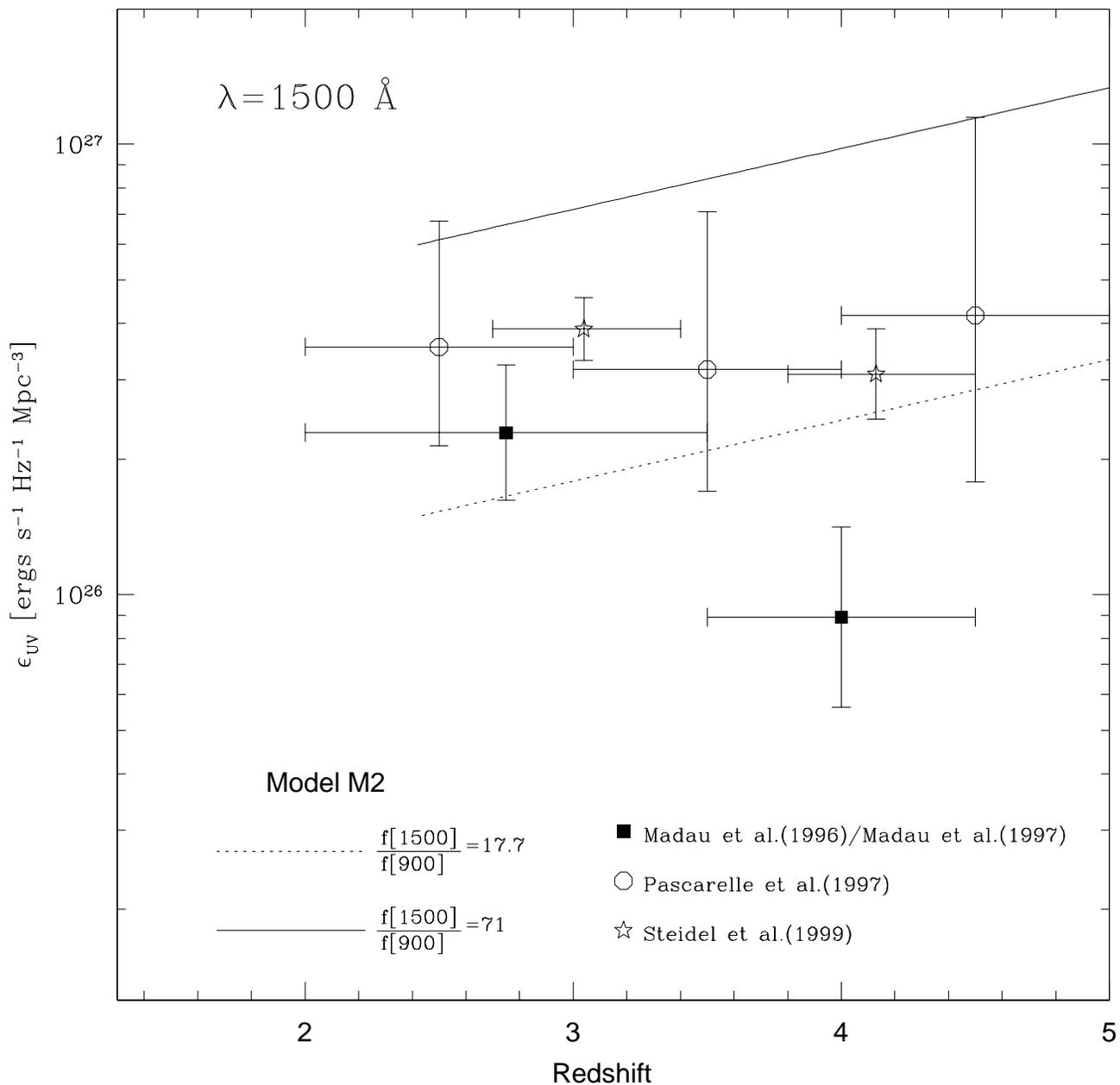}}
\end{picture}
\caption{
Comoving emissivity at $\lambda=1500$ \AA \ from our
galactic Model M2 as a function of redshift based on the two values of
the observed flux ratio from Steidel et al. (2001) (dotted line) and
Giallongo et al. (2002) (solid line).  Also plotted are the relevant
observations of the UV emissivity at $\lambda=1500$ \AA \ from Steidel
et al. (1999), Pascarelle et al. (1998), Madau et al. (1996), and Madau
et al. (1998) adjusted for a flat $\Lambda$ cosmology: ($\Omega_m$,
$\Omega_{\Lambda}$)=(0.37, 0.63) and $h=0.7$ (corrected data obtained
from Nagamine, Cen, \& Ostriker [2000]).}
\end{figure}

\begin{figure}[htb]
\figurenum{6}
\setlength{\unitlength}{1in}
\begin{picture}(6,6.5)
\put(-0.65,-1.7){\includegraphics{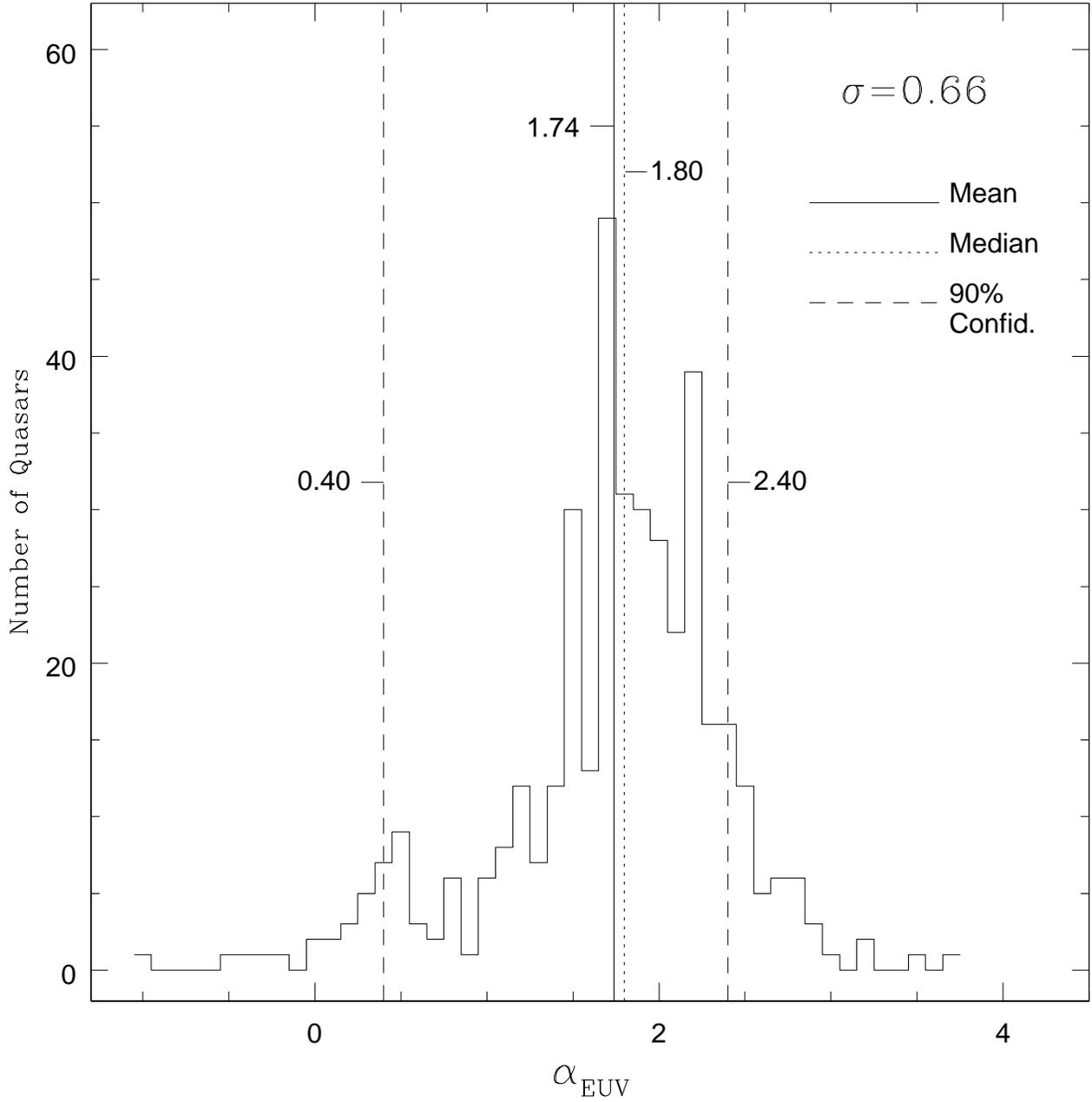}}
\end{picture}
\caption{Quasar EUV spectral index distribution compiled from the data
in Telfer et al. (2002). The distribution includes results for
subsamples of radio-quiet and radio-loud QSOs (see text). The indices
are defined as $f(\nu)\propto\nu^{-\alpha}$ between 500 and 1200
\AA.  The mean, median, and rms deviation ($\sigma$) of the
distribution are indicated.}
\end{figure}

\begin{figure}[htb]
\figurenum{7}
\setlength{\unitlength}{1in}
\begin{picture}(6,4.5)
\put(-0.65,-1.7){\includegraphics{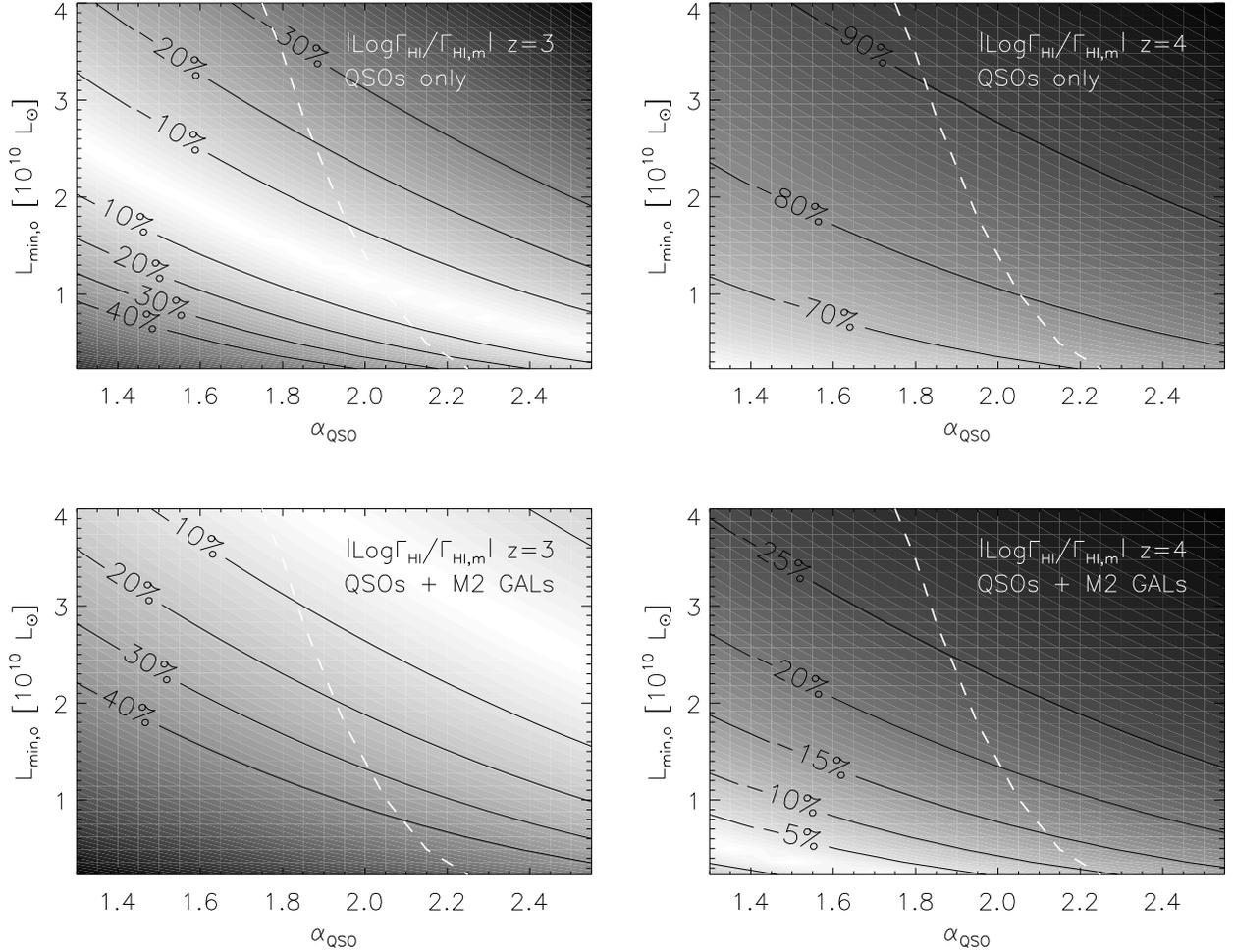}}
\end{picture}
\caption{Greyscale plots showing the absolute value of the difference
between the logarithms of the computed hydrogen photoionisation rate,
$\Gamma_{\hh}$, and the measured value obtained from Rauch et
al. (1997), $\Gamma_{\hh,m}$, as a function of quasar parameters
$L_{\rm min,0}$ and $\alpha_{\rm QSO}$ at redshifts 3 and 4. The top panels
show the results for the case with only quasars while the bottom
panels show the results for quasars with M2 galaxies included. The
labeled contours in each panel show the percent difference from the
corresponding measured value. Also plotted in each panel is the curve
({\it white-dashed line}) in the $L_{\rm min,0}$-$\alpha_{\rm QSO}$ 
plane that
would produce the same number of He {\small II} ionising photons as
quasar Model 5 from SAH.}
\end{figure}

\end{document}